\documentclass[12pt]{article}

\begin{document}

\author{\textbf{Heather M. Roff}\\Senior Research Analyst, Johns Hopkins Applied Physics
Laboratory\\Fellow, Foreign Policy, Brookings Institution\\Associate Research
 Fellow, Leverhulme Centre for the Future of Intelligence,\\University of
 Cambridge}
\title{\textbf{Expected Utilitarianism}}
\date{ }   
\maketitle

\section*{Introduction}
We want artificial intelligence (AI) to be
beneficial.\footnote{https://futureoflife.org/bai-2017/}

This is the grounding assumption of most of the attitudes towards AI research.
We want AI to be ``good'' for humanity. We want it to help, not hinder, humans.
Yet what exactly this entails in theory and in practice is not immediately
apparent.
Theoretically, this declarative statement subtly implies a commitment to a
consequentialist ethics.
Practically, some of the more promising machine learning techniques to create a
robust AI, and perhaps even an artificial general intelligence (AGI) also commit
one to a form of utilitarianism. In both dimensions, the logic of the beneficial
AI movement may not in fact create ``beneficial AI'' in either narrow
applications or in the form of AGI if the ethical assumptions are not made
explicit and clear.

Additionally, as it is likely that reinforcement learning (RL) will be an
important technique for machine learning in this area, it is also important to
interrogate how RL smuggles in a particular type of consequentialist reasoning
into the AI: particularly, a brute form of hedonistic act utilitarianism.
Since the mathematical logic commits one to a maximization function, the result
is that an AI will inevitably be seeking more and more rewards.  We have two
conclusions that arise from this. First, is that if one believes that a
beneficial AI is an ethical AI, then one is committed to a framework that
posits `benefit' is tantamount to the greatest good for the greatest number.
Second, if the AI relies on RL, then the way it reasons about itself, the
environment, and other agents, will be through an act utilitarian morality.
This proposition may, or may not, in fact be \emph{actually} beneficial for
humanity, (at least for the majority of reasons cited against utilitarianism
over the past three hundred years). Indeed, as I will attempt to show here,
much of the philosophical insights about the deficiencies of utilitarianism
apply directly to many of the currently cited concrete problems in AI safety,
such as specifying the wrong objective function, avoiding negative
externalities, reward hacking, distribution problems, epistemological
limitations, and distributional shift\footnote{https://arxiv.org/abs/1606.06565}.

The paper is organized into four sections. The first section lays out how RL
could be seen as an implementation of hedonistic act utilitarianism. Section
two discusses the various kinds of utilitarianism, as well as many of its
well-cited criticisms, particularly how most of its forms ultimately collapse
back into a brute act utilitarian evaluation. Moreover, this section argues
that the two ways that utilitarianism is often described, as either an
evaluative theory of right or as a decision-procedure for individual action, is
the same description of how machine learning researchers are approaching the
problem of creating beneficial AI. This presents us with a useful intervention
point, both within policy but also within research communities to assess
whether this is a solvable problem for AI. The third section argues that many
of the problems noted in classic arguments about utilitarianism manifest
themselves in concrete AI research safety problems, especially in reward
hacking. Finally, section four argues that it may be useful for AI researchers
to look towards philosophers to see how they have previously attempted to solve
such problems. In particular, it may be more simple to look to various
constraints on an AI's action, or as Mackie notes as ``device[s] for countering
such specific evils'' that may result\footnote{Mackie, J.L.
\emph{Ethics: Inventing Right and Wrong} (Penguin Books, 1977): 139.}.

\section{Reinforcement Learners or Hedonistic Act Utilitarians}

The typical reinforcement learning paradigm divides any problem into three
parts: a policy; a reward signal; and a value function. Some, such as Sutton,
also include an internal model of the
environment\footnote{http://www.cell.com/trends/cognitive-sciences/fulltext/S1364-6613(99)01331-5}.
The policy maps states to actions or choices; the reward function signals to
the agent what is immediately ``good'' given the present state and choice of
action, and the value function allows an agent to evaluate its present state
in relation to some future.
The internal model of the environment is however slightly trickier from a
conceptual and ontological viewpoint, and as many researchers would argue is
not necessary for an RL agent. Perhaps it is mere semantics, but there also
seems to be a need for an agent to represent various information inputs from
the environment, say through sensory perception or other signals, to create its
own ``internal'' model of said external environment. Through iterated and
continuous experiences, the agent can build a more robust model of it, or
perhaps merely of states likely to give it more rewards in the future.

Additionally, there is a background assumption that the agent and the
environment are external to one another. In short, from a conceptual and
mathematical perspective, agents do not help to construct part of the
environment in which they act.
We can view this as a \emph{billiard ball model},\footnote{See for instance
Kenneth Waltz, \emph{A Theory of International Politics} (Waveland Press,
1979), where he describes 2. “Structural Realism” as a billiard ball model
with states as the billiard balls and the international system as the table.}
where the environment, i.e. the table (and perhaps other agents-as-billiard
balls), structures the experiences and possible configurations of actions for
an agent, but remains distinct from the agent itself.

This assumption is crucial for at least two reasons. First, it denies a
``constructivist model'' for learning. Constructivism as a social theory,
claims that agents and structures are not ontologically separate, but that
they co-constitute each other. While the theory is primarily concerned with
the behavior of agents with other agents, and as such potentially may have
some limitations to single agent worlds, it denies that an agent does not
change its environment through its action and vice versa. The environment can
redefine an agent through time, and an agent can redefine its environment;
neither are static.\footnote{See Alexander Wendt, \emph{Social Theory of International Politics} (Cambridge University Press, 1999) and Nicholas Onuf “A Constructivist Manifesto” in Constituting International Political Economy, Eds. Kurt Burch and Robert A. Denemark (Lynne Rienner Publishers, 1997): 7-20.} A constructivist learning model would grant that an agent may change its environment --- its inputs or signals or structuring conditions --- and that these factors also change the ``internal'' model of the agent.

While RL is premised on a feedback loop, where an agent tries a particular
action or behavior to reach a (preselected/identified) goal and receives a
reward signal from the environment providing information about whether it is
acting correctly, the environment does not \emph{fundamentally} change. The
environment may produce varying signals to the agent, but the environment as
such, does not ontologically change, and so this is not a co-constituting
ontology. The feedback loop only goes in one direction, not two. Thus, for the
RL agent, a goal remains something external or exogenous to it, and the agent
directs its actions towards it. It cannot modify or change its goal. This is
important because as AIs become more powerful, the model for learning may need
to change. RL agents may be able to reshape not only their environments, but
rewards as well.

An example may be helpful here. An agent playing chess receives various
signals from its environment (such as another agent's move on the board). The
signal tells the agent which actions are not only permissible to take, but
which actions may preferable or best to take, given the available knowledge
about the game of chess at the time, with the goal of winning the game. This
is a feedback loop. However, it is not a co-constituting agent-structure
problem. The agent playing chess cannot, once its adversary moves to e5, claim
that the board itself changed and now is a 9x9 matrix. The board is the
structure; the agents move within it.

The conceptual terrain is important here to understanding how the logic of
reinforcement learning and the billiard ball model of learning work to produce
a particular kind of utility maximizing agent: a hedonistic act utilitarian.
While one may object, and claim that the RL utility maximizing agent is a
different kind of ``utilitarian'' than that of the moral philosophers, this is
in fact, not wholly true. \emph{One is merely the mathematical representation
of the logic of the other}, and indeed implicitly carries in certain
assumptions about broader normative aims.

For example, as Sutton and Barto (1998) explain with regards to reinforcement
learning,
\begin{quotation}
Rewards are in a sense primary, whereas values, as predictions of rewards, are
secondary. Without rewards there could be no values, and the only purpose of
estimating values is to achieve more reward.

Nevertheless, it is values with which we are more concerned when making and
evaluating decisions. Action choices are made based on \emph{value
judgments}.\footnote{Sutton and Barto. \emph{Reinforcement Learning}
(1998): 8.}
\end{quotation}

Here, a value judgment is a judgment about what sorts of behaviors or actions
contain ``worth''. But what ``worth'' entails, is actually about what is
``good'' all things considered for the decision or task at
hand.\footnote{One can actually discuss two different kinds of worth here. The
first is merely what the agent will likely choose in the present state taken
discounted future states and choose the one that maximizes. The other is to
look from one step back and claim that how one specifies the reward
function--as a signifier of what is desired/right/correct--is indicative of
some sort of behavior that possesses \emph{worth} (or utility).
In this sense, the objective (that is the goal) is that which has worth, and
thus the objective function for a given task enumerates the possible values
for various state and action combinations.}
It is, therefore a teleological or goal-directed action. Moreover, if one
desires to avoid circularity, then one cannot define ``good'' or ``value'' in
terms of itself, so one must find some other thing to use as a signal or
content for the ``good.''
In the RL scheme, the signal is the reward function, and this is used to
estimate or ``judge'' which actions or states will yield the long term best
act/choice/decision; that is, what will most likely yield the ``value'' function.

Yet if we grant that this conceptualization of RL as true, then we must also
grant that it maps --- almost identically --- to the logic of the moral theory
of utilitarianism as well. Utilitarianism is a \emph{consequentialist moral
theory} that argues that normative properties are determined by some sort of
end state (or goal). In easy terms, the best consequences --- here
operationalized as the most utility --- determine the morally right act.
Utility can be defined in any number of ways, as I will explain in the next
section, however it is most often associated with an aggregate welfare
account, usually seen in shorthand as ``the greatest good for the greatest number.''

One ought to immediately see that RL is very much like utilitarianism because
both the RL agent and the utilitarian moral agent seek to determine some
present action by a judgment about maximizing the value --- or good --- of
some future state/goal/consequence. Additionally, utilitarianism also seeks
coherence and avoiding circularity. Thus, one cannot define the good in terms
of itself, but must instead define it in some other (nonmoral) term.
Classically, this is done through a reward signal (pleasure or
pain).\footnote{Cf. Jeremey Bentham. An Introduction to the Principles of
Morals and Legislation, 1789. (Prometheus Books, 1988). John Stuart Mill.
\emph{Utilitarianism}. Henry Sidgwick. \emph{Methods of Ethics}, seventh edition (MacMillan 1907).}
The utilitarian uses the reward signal to determine which act is the right
one. For instance, given options x, y, and z, she determines that action y
will give her the most pleasure or happiness, and so she chooses to follow
this. She uses the feedback of pleasure and the instruction to maximize it as
the decision criteria to judge which of the actions to take. Her pleasure,
seen as akin to happiness or an indicator of the value of happiness, is the
signal that tells her what she ought to do. Likewise, her pain tells her what
to avoid.

The logic is the same in both accounts. First, because utilitarianism as a
moral theory, takes as its starting point the position that there is an
\emph{objectively right answer} (\emph{a la} moral realism).
Utilitarianism is not a doctrine of egoism, but instead claims that there is
some external good that exists external to the agent, and that one ought to
comport one's actions with this good.
It is thus ontologically similar to RL in that it takes the environment as
separate from the agent, i.e. that there is some objectively distinct
environment. Thus the definition of the reward must be external and outside
the control of the agent too.

Second, both RL and utilitarianism are goal seeking and reward maximizing. RL
agents seek to maximize their rewards, either in the immediate or discounted
for long-term gains, given an uncertain environment and limited knowledge (in
philosophical terms ``epistemic constraints''). As Sutton and Barto explain, ``a reinforcement learning agent's sole objective is to maximize the total reward it receives in the long run.''\footnote{Sutton and Barton, op. cit., 8.} In both accounts there is ambiguity and contestation about what the
appropriate time horizon ought to be.
The ``right'' or ``best'' decision is heavily dependent on the goal or end
state. Additionally, we would perhaps want to claim that the immediate goal is
reward maximizing, but this is only coherent when we view it from the
perspective of a secondary goal (defined by the objective or task of the
agent). Utilitarianism's logic is also goal seeking and reward maximizing. The
agent is told to maximize utility in everyday decisions, and that this will
achieve a secondary goal of increasing the overall utility for all agents
``in the long run.''

Third, on first glance RL policies seem to mimic much of the logic about
utilitarianism \emph{as a decision procedure}. Recall that a policy simply
maps an agent's perceived states about the external environment to the menu of
potential actions available to the agent. The agent then may use its prior
experiences of the reward signal, as well as its vague knowledge about the
value function, to make a decision. The policy is, then, best represented as
the result of a decision-procedure based on prior experience and a present
directive.

Utilitarianism, likewise, is often touted as a decision-procedure. While many
try to separate this from other versions, such as
utilitarianism-as-a-criterion-of-rightness, often it collapses into a mere
decision procedure. This is best represented in its ``act utilitarian'' form.
That is, every act A in situation S must be evaluated on whether or not it
will maximize utility U, where U is operationalized as some nonmoral good
(happiness, pleasure, welfare, etc.) compared to all other available acts. Act
utilitarianism, however, yields highly unstable and often counterintuitive or
undesirable results when taken in the aggregate or under long-term
considerations. Agents may be individually acting ``rightly'' but collectively
they are not. For example, I might consider that it will benefit me more to
cut in front of a person in the queue. I feel that my time is worth more, and
it will bring me more pleasure to ignore the other person's place. However, if
all people began acting like this there would no longer be queues, and indeed there would most likely be bedlam.

Utilitarianism-as-a-criterion-of-rightness, however, seeks a particular
``indirect'' form of utilitarianism that does not utilize a direct maximizing
procedure (such as a greedy algorithm). The right act is right according to
the principles of utilitarianism in that the act does maximize utility, but
how one decided to pursue that act, such as by a rule, law or conventional
norm, is perfectly open. Unfortunately, this version of utilitarianism, is
often charged as being self-defeating because it collapses back into the
cruder version of act-utilitarianism when pressed, such as with the case of
rule-utilitarianism.\footnote{Kymlicka, Will. \_\_\_. Brandt, R.S.}

Rule-utilitarianism claims that in situation S, an agent ought to follow Rule
R, and that the agent should not attempt to calculate the various utilities of
all available actions to see which act \emph{actually} maximizes U. Rather,
the Rule R, ought to be followed because the general observance of R will on
par maximize utility for the aggregate over time, even though individual
agents may fail to individually maximize their respective utilities through
their individual acts. This is a sort of long-term vs. short-term, individual
vs. society, balancing intended to get around the various perverse outcomes of
direct act utilitarianism. Unfortunately, the rule-utilitarian faces charges
of ``rule worship'' when they find themselves thinking that a right act in S
actually requires them to violate the rule.

With respect to RL, both RL agents and moral agents are in the same logical
position when it comes to deciding which action it/she ought to undertake. In
short, both the RL agent and the moral agent attempt to figure out or to
determine the right act based on some non-moral signal about what the ``good''
is, and what is more, both are faced with the theoretical problems of
disentangling the criterion of rightness from the immediate decision
procedure. \footnote{I should note that presently the RL agents may not be
capable of this, but as they become more sophisticated this problem will arise
if the logic remains the same.} Thus contra Soares' and Fallenstein's
assertion that philosophers do not provide descriptions of ``idealized
decision procedures'' in their work on decision making, it is in fact the
primary focus of much of the work on justifying utilitarianism as a moral
theory, both in idealized form and as an action-guiding theory (\emph{qua} decision procedure).\footnote{Soares, Nate and Benja Fallenstein. “Toward Idealized Decision Theory” unpublished essay. arXiv:1507.01986v1 [cs.AI] 7 July 2015.}

Since there are problems with splitting utilitarianism into two forms, that of
utilitarianism-as-criterion-of-rightness and
utilitarianism-as-decision-procedure, it seems that the most coherent
framework is actually to utilize them both simultaneously. This logic needs to
link the end value of utility (as defined by X) to the immediate decision
through some sort of estimated calculation and the accompanying nonmoral
signal (pleasure, pain, reward, punishment). Trying to decouple these concepts
leads to at least to inconsistencies and, at most, to violations of the
founding principles of utilitarianism (and probably to why we even design AIs
to aid in our daily life).

Moreover, if one grants that RL agents follow the same logic as I've outlined
above, particularly when it comes to using the immediate reward signal as a
learning or communicating device about the value function, then what we must
accept is that this logic entails that an RL agent is akin to a hedonistic act
utilitarian. Whatever the ``reward'' is, it is something the RL agent pursues
through maximizing. Additionally, because the logic of both act
utilitarianism and RL works the same, one can see how the under-specification
of ``the good'' and/or the ``value function'' or ``objective function,''
as well as how this is coupled with a reliance on some nonmoral immediate
reward signal, leads to unintended and undesired outcomes. In the case of
ethics, we may reach ``repugnant conclusions,'' and in the case of RL,
problems of AI safety. In either instance, this is due to the internal logic
of the theory.

\section{Utilitarianisms \& the potential for interventions}
We have briefly discussed two of the most common forms of utilitarianism in
the previous section, act utilitarianism and rule utilitarianism. However,
over the last hundred years there have been other attempts to overcome the
previously noted problems, e.g. self-defeating, over-demandingness, perverse
outcomes, by introducing other forms of, or amendments to, utilitarianism. In
essence, the strategy for solving some of the undesired outcomes of a straight
maximizing hedonistic act utilitarian is to make explicit, or to put into
practice, how one goes about maximizing utility. To this end, we need a
definition or content of utility and a rule for counting (and thus maximizing).

Defining what constitutes utility --- beyond hedonistic conceptions --- has
taken the forms of: mental state utility, or having the mere belief or
experience of pleasure or happiness; preference satisfaction, which could be
adaptive preferences (those that change), informed preferences (such as from a
rationalist or ``external'' standpoint), or personal preferences, that may or
may not be irrational or selfish.\footnote{CITATION.}
We can also couch these forms of utility in broader categories, such as
political utilitarianism, which looks to provide utility in major social
institutions, or solely to comprehensive moral utilitarianism that regulates
the right conduct between moral agents.\footnote{..}
The simple point is that any definition of utility needs to provide the
\emph{type of value} that an agent ought to seek. For purposes of RL, we can
see that it is the way that the reward function is designed that will
ultimately comport with, realize or undermine whatever normative value we
place on the tasks or goals we desire the agent to accomplish.

Additionally, utilitarians make choices about where one ought to view the
\emph{locus of value}. This can be examined in two ways, either as: agent
neutral or agent relative. For an agent neutral perspective, the utilitarian
agent does not perceive value from her perspective, but ought to do so from an
impartial observer's standpoint. For an agent relative point of view, what
counts as ``good'' is determined by the point of view of the agent. These two
distinctions are important because they permit or deny us to weigh the
interests of others with greater or lesser degree.

For instance, if I value the interests of my fellow citizens, friends or
family more, and give them greater weight in my calculation, then I will get a
far different outcome than if I weigh each individual's interests without
reference to their relationships to me. In some cases, we may think this is
entirely beneficial, but in others less so. Take, for example, the common view
that we do have associative duties to particular people, such as parents to
children, or spouses or particular role responsibilities; if this is true then
we cannot explain or justify those under an agent neutral utilitarian
perspective.\footnote{Lazar, Seth.} However, an agent neutral perspective does
tend to guard against particular preferences that we think ought not to be
counted, such as those that may undermine the well-being, rights or lives of
others.

An additional view looks to challenge the notion that there is one singular
value to be maximized at all. Rather, this perspective supports the position
that there is a \emph{plurality of values}.\footnote{.. } Due to this plurality, however, there are often questions related to incommensurate or conflicting values, or even of ordinal or lexical rankings of them. Thus this perspective requires a corresponding theory of distribution, by which goods, interests, values, etc. are ranked and meted out.\footnote{.. }
This requirement must then deal with issues of fairness, desert, and
trade-offs, and for the theory to be coherent, there must be a consistent and
nonarbitrary way of formalizing these rules or constraints. A theory of
distribution, then, is tightly linked to utilitarianism's requirement to
provide some sort of rule for counting (and thus maximizing)
utility.\footnote{Kymlicka.}

For RL, however, one might argue that one does not need to specify any of
these elements. The RL agent will learn them, given enough data, experience
or episodes. This, however, is not necessarily true, and as I argue in the
next section, there are specific design choices that may construct or
determine these judgments without the designer's conscious choice.

For example, as Kymlicka explains with regard to utilitarianism, philosophers
view the maximizing rule from two different perspectives: the egalitarian view
and the teleological view. The egalitarian espouses that maximizing is a
\emph{byproduct} or effect of utilitarianism. In essence, it takes the rule to
maximize as a derivative, indirect and necessary feature of the theory.
Maximization is not a primary and direct goal for an agent to realize, rather
it is just ontologically required. The teleological account, contrarily, takes
maximizing as \emph{the goal state}. Maximizing the good is both using
maximizing as the decision procedure as well as the criterion of rightness.
Immediately we can see that for the RL agent, maximizing is not seen as a
byproduct of an agent's actions, but is, both from a formal mathematical and
theoretical perspective its goal state. While the immediate reward will affect
an RL agent's policy and thus choice in action, the long-term goal is still to
maximize rewards as much as possible. Moreover, whatever the objective
function is for an agent, there is some implicit acknowledgement that it is:
(a) good all things considered, and (b) that one ought to maximize it.

Utilitarians, moreover, need to provide further particulars about the rules
for counting. While it is true that in some (direct or indirect) way
maximizing utility will occur, we still need to figure out \emph{who} is in
the population of individuals to be counted and as well as \emph{what} sorts
of things we count or discount. For instance, the brute aggregative style of
utilitarianism says that the total utility is the sum of its individual parts,
whereas a ``total'' form of it may say that it is not aggregation but a net
good (however that is determined).\footnote{..} Or some, like Nick Bostrom,
would take a universalist conception of utility where all sentient beings in
the past, present and future states must be counted for the total benefit of
``humanity.''\footnote{..} Or even others expand the possible pool from not merely generations, but also the kinds of good to be included, such as acts, rules, motives, or even sanctions.\footnote{..}

Moreover at work in many of these counting exercises is, unfortunately, a
privilege of quantity over quality. As Parfit notes, there is a paradox -- the
``mere addition paradox'' -- that some state of affairs may be logically more
desirable because there are a larger number of people who have some, very
minimal, life ``barely worth living,'' than a different state of affairs where
there are fewer people but who all have a very high quality of life. The mere
addition of more destitute and desperate people is judged to be morally
preferable over the addition of fewer happy and secure individuals. The
paradox results in what Parfit calls the ``repugnant conclusion'' where most
people would conclude that the first state affairs is not in fact more
desirable, let alone acceptable. (A sort of: Maths do not morals make.)

Again we can see that decisions about the design of the reward signal are
paramount, for how the agent is rewarded in particular states will determine
which choices it makes. I would submit, moreover, that how the environment is
designed, as well what data is provided, will also color how and what agents
learn. For instance, if we claim that an optimal policy is \fbox{$\prod$}, then, we have \emph{a priori} determined what the optimal value for an agent is in this
particular instance (state) for this particular decision
(act).\footnote{Or that the agent has found the policy itself if one prefers that language. The important point is that the design of the reward is a normative choice.} To \fbox{$\prod$} may be to count each individual's interests
equally, with no regard for associative duties or bonds, or to not consider
the quality--rather than the quantity--of the good of each individual. In
which case, \fbox{$\prod$} would look like Parfit’s repugnant conclusion.

Additionally, utilitarianism has grappled with the notion of whether agents
ought to calculate the actual consequences of their actions or the expected
values of future states. While the theory appears to posit -- as a criterion
of rightness -- that it is the actual consequences that matter, others argue
that it is instead the foreseeable or expected values associated with likely
future consequences that matter.\footnote{..} Humans cannot truly calculate
utilities before every decision, as they are imperfectly rational and live
under conditions of uncertainty. However, they can, given the right
information, make predictions about the expected outcomes of particular
choices. Thus, we ought to make moral judgments on the foreseeable outcomes of
decisions and actions, not on the actual ones.

This perspective also comports, both conceptually and mathematically, with
much of the work in RL. For, if it is true that the right act (or perhaps
policy) is that which maximizes utility, then as I've argued it is right both
in the immediate case and is used as a formal decision procedure, but also it
is right in the normative teleological case (the value function). The goal,
and the value of reaching this goal, determine which acts are right in
relation to it. However, because an RL agent is always calculating the
probability of maximizing its reward over time, it is calculating its current
expected utility and the estimated values of future states of action (and in
some cases bootstrapping the current estimate to the estimate of successor
states). We can update our value functions when the actual circumstances come
to pass, thereby improving our policies, but we do so on a continual feedback
between estimates and future acts.

Ultimately, what is often overlooked, is that two ways that utilitarianism is
described, as an evaluative theory of right and as a decision-procedure for
individual action, is the same description of how machine learning researchers
are approaching the problem of creating beneficial AI and, potentially, AGI
from an RL methodology. Beneficial AI, through RL, is that which provides for
the many over the few, generates more welfare rather than harm, and even
``aligns'' with human values\footnote{CITATION}. Additionally implicit is that
these types of values ought to be maximized to the greatest possible extent,
for our ``cosmic endowment'' as humans. This way of viewing AI- from a
normative perspective - is then mirrored in the way RL actually works: through
the maximizing of rewards to learn a policy.

We must be careful, then, when discussing beneficial AI, or perhaps more
expressly ``moral AI,'' because we begin to blend normative and empirical
facts together. Since both utilitarianism and RL share a common logic, we can
take this opportunity as a useful intervention point, both within policy but
also within research communities, to assess how to create beneficial AI. Many
of utilitarianism's problems relate the strict instruction to maximize utility
despite the complexity of the world, where humans have various commitments, a
plurality of values, and other contingent facts. Beneficial AI based on RL,
without careful attention, faces the same sorts of struggles.

\section{AI Safety Problems \& The Pitfalls of Utilitarianism}
To be sure, it almost seems a truism to claim that we ought to have beneficial
AI, for no one is claiming that we ought to have harmful AI. Rather, it
appears that most concerns surrounding safety pertain to mitigating or
eliminating potential harm from AI on a negligence basis. Indeed, recently
Amodei \emph{et al}, identified several current and pressing `Concrete
problems in AI safety' regarding the potential for harm or imposition of risk
of foreseeable AI-related harm. Specifically, they note three classes of
potential areas for risk: specifying the wrong objective function; possessing
an objective function that is too expensive to evaluate frequently; limiting
undesirable behavior during the learning process.\footnote{Amodei, Dario,
Chris Olah, Jacob Steinhardt, Paul Christiano, John Schulman, Dan Mane. ``Concrete Problems in AI Safety'' 2016.}

These are then refined further into specific problem types, all of which have
to do with not only the formalization of the task but also how the agent's
learning or completing of this task is nested in a wider socio-technical
system of values. While Amodei \emph{et al} identify the difficulty of correctly designing and engineering AI agents, they also implicitly acknowledge that these difficulties are not solely technical. Rather they are
``risky'' or ``harmful'' because AI agents are increasingly intertwined with
environments that often include or affect humans and human values.

Interestingly, many of these problems, such as avoiding negative side effects
and avoiding reward hacking are just new manifestations of some of the oldest
problems of utilitarianism. To see why this is so, let us briefly take each in turn.  First, the problem of negative side effects: as Amodei \emph{et al} explain, ``for an agent operating in a large, multifaceted environment, an objective function that focuses on only one aspect of the environment may implicitly express indifference over other aspects of the environment.'\footnote{Ibid, 4.} In plain terms, we may create agents that are
very good at doing one thing but lack the capability to understand or estimate
how their actions related to that one thing affect a wider community of
agents, interests or values.

Here we may claim that the task and the objective function for that task might
be poorly or underspecified, and that there is some sort of negligence at work
when a designer did not reasonably foresee that the agent would act in this
way. However, this is not really an objection to RL, rather it is presented as
merely one of the difficulties of creating sophisticated RL agents.

Utilitarianism struggles with the same difficulty: creating moral agents who
do not undermine the rights (or values, or interests, etc.) of others while
pursuing their endeavors. Like the RL agent, the utilitarian is told to act so
that she achieves the greatest good (highest reward), and as I have suggested,
she is also given non-moral signals as to what that ``good'' is (through
feelings of pleasure and avoidance of pain). As Mill explains,
\begin{quotation}
As I see it, then, for me to have a right is for me to have something that
society ought to defend me in the possession of. Why ought it to do so? The
only reason I can give is \emph{general utility}. If that phrase doesn't seem
to convey a good enough sense of how strong the obligation is -- i.e. good
enough to account for the special energy of the feeling -- that's because the
make-up of the sentiment includes not only a rational element but also an
animal element [$\ldots$].\footnote{Mill, John Stuart. Utilitarianism}
\end{quotation}

Unfortunately, the logic of this viewpoint is to ground all rights in social
acceptance of a thing that the majority of people find important, and to then
base this feeling of importance on one's strength of feeling. Thus depending
upon one's society, values, education and views of those outside of one's
society or ``different'' than the majority, the notion of having a right
begins to lose a lot of weight. Rights are not ``trumps'' against the weight
of opinion, but are only accorded through and by that opinion. Ultimately the
utilitarian admits of the need of three other contingent facts about humanity:
capacity for moral sentiments or feelings; education; sociability. This is
where RL agents and utilitarians begin to diverge slightly.

Where the RL agent ``finds'' the best policy, the agent has little
understanding or guidance on the overall goal of the system. In this way, RL
agents can quickly and uniquely find correlations between various types of
actions and states and the rewards that the environment signals to them.
However, in many instances these correlations between signal and learning will
lead to spurious results because the agent does not have reference to the
goal. It seeks to maximize its reward function not in any teleological sense,
but purely by trial and error.

Utilitarians, moreover, would also suggest that trial and error are
fundamentally important in learning to be a virtuous moral agent. However,
this trial and error must be grounded in human ontology, both at the
individual and societal levels, and not from a view from nowhere. Human
affairs are necessarily social, and if one believes Locke, even the family is
seen as the ``first society.''\footnote{Locke, John. ``The Social Contract''}
The difficulty for utilitarianism then is to resolve the tension between
utilizing the individual's reward signal (pleasure/pain) to teach or guide
action for her, while simultaneously broadening that signal to include a
variety of agents and goods (others, society, etc). In the local and immediate
sense, pleasure or pain works very well to signal to a person to pursue or
avoid certain things. However, using this same metric to guide action at scale
for ``society'' or ``humanity'' becomes more difficult. In this way,
utilitarians rely on systems of moral education but also on the ``firm
foundation'' of ``the social feelings of mankind'' to argue that the ``social
state'' is a natural one, and that by not taking a utilitarian stance
(greatest good for the greatest number) is unnatural.\footnote{Mill,
John Stuart. Utilitarianism, 45. Kant likewise refers to this sort of scheme,
though he likes to claim that humans are ``unsocially sociable'' in that they
want to pursue their own projects without interference from others. Cf. Kant,
Immanuel. ``Idea of a Cosmopolitan History with a Universal Purpose.''}

In the plainest terms, utilitarians see the world as including a myriad of
values and interests, all of which are connected to other agents. My pursuit
of one pleasure cannot ``deprive mankind of the good, and inflict upon them
the evil'' of my action.\footnote{Mill, op. Cit., 34.} The utilitarian agent
must use her natural capacities for feeling but also gain education and
training in how to recognize and balance one particular project against many,
one subgoal for a higher-order one, and quality as much as quantity. In short,
she cannot be myopic or a savant.

This vision of the utilitarian agent, however, plausible on its surface faces
many difficulties when put into practice. Calculating utilities can include
too much, and be over demanding for an agent, or include too little and
violate our common intuitions about what behavior is morally acceptable. It
does not tell us if the act that we are undertaking does in fact undermine or
detract from the greatest good. It does not really tell us in any specific
terms what the ``good'' for society actually is. In effect, the moral theory
doesn't give us good guidance on how to avoid negative side effects.

The RL agent, however, has no moral sentiments, and indeed it is not
\emph{a priori} social (nor even multi-agent). All it has is the reward signal
and the instruction to maximize. Seen in this way, it is at an even further
disadvantage than the utilitarian moral agent in being able to deconflict its
actions from potentially negative side effects. The moral agent at least has
recourse to some notion of a goal, however abstract, as well as some notion of
competing objectives.

One might object here and claim the RL agent is capable of doing such things,
and in fact, there are two ways around this. First, we could give an RL agent
multiple objectives thereby trying to over-determine the “right” act for the
agent. Second, we might not need to determine the objective function at all.
We might leave it unspecified and let the agent work it out on its
own.\footnote{For one example, we could do some form of inverse reinforcement
learning.} While each of these are certainly methodologically true, there is a
deeper flaw in this logic. First, even if we could give an agent multiple
objectives, this does not provide any principled way for the agent to either
non-arbitrarily choose amongst them and it admits of a fact-value
problem.\footnote{In cases of competing or conflicting objectives the tendency
seems to be to likewise institute some arbitrary way of deciding amongst the
objectives (e.g. lexicographically, linear preferences or pareto dominance).
Merely averaging them may be useful and simple, but that does not get us out
of the negative side effects problem.} Second, even if the RL agent works out
the objective function on its own, it too still suffers from a fact-value
problem. 

Fact-value problems, or naturalistic fallacies, are problems of logic. In
short, as Hume noted in 1739, one cannot derive an ``ought'' from an ``is.''
Just because the world is a particular way (fact) does not mean that people
ought to act a particular way (value). It basically is to conflate moral facts
or properties with natural ones. Thus we might claim that from all the reward
signals an RL agent receives it figures out the optimal policy, but that
policy does not carry with it any moral weight. One cannot admit of any moral
prescription or imperative based on them. Just because the agent figures out
an objective function does not mean that it ought to act according to it. Thus
even if it ``aligns'' its value function with what I value, this does not in
any way speak to its moral praiseworthiness.

The ``avoiding negative side effects'' argument then tends to look like one
where the RL agent acts in ways that I want it to act, and not to be harmful
to my interests. This is deemed ``beneficial.'' Yet when pressed, this notion
of beneficial requires of agents that they know my multiple values, and I
suggest, a way of knowing which ones to privilege and which ones to discount.
I might, for instance, be a misanthrope who secretly desires to go about
pinching people for my pleasure. The RL agent, it would seem, ought not to
provide me with detailed plans about how best to realizing my pinching
pleasures. The agent would thus need to know my value set and the ``moral''
value set and be able to arbitrate between the two to truly avoid such
negative side effects.

Second, reward hacking is also just old wine in a new bottle. Indeed, we can
predict that RL agents under a maximizing directive will -- just like
utilitarian agents -- ``reward hack.'' Why is this so? As viewed from a human
designer's perspective, the agent, because it is reward seeking and
maximizing, finds a way to gain more reward in a formally valid way, but by
doing so is acting outside of the bounds of designer's intent. The more
sophisticated the agent becomes, the more complex the environment and task,
the more likely it is that the agent will find some way to find a solution not
foreseen by the designer to achieve more reward without, or perhaps not
sufficiently, accomplishing or satisfying its tasks.

We call this a ``hack,'' such as some version of an ``exploit,'' and thus it
carries with it a negative connotation. However, it really isn't negative from
an agent's perspective. It is completely permissible and encouraged given the
requirement to maximize. The negativity only arises in relation to some
external set of value judgements, such as what the designer wants or does not
want it to do.

For the utilitarian, we don't call it a reward hack, we instead are faced with discussing ``the life worth living.'' For instance, in one of his most famous
quips, Bentham acknowledges that it does not matter if one favors ``pushpin to
poetry;'' they are of equal moral worth to a utility maximizing
agent.\footnote{Jeremy Bentham, from \emph{The Rationale of Reward}, excerpted
and reprinted in: \emph{The Classical Utilitarians: Bentham and Mill}, ed.
John Troyer (Indianapolis: Hackett, 2003), p. 94.} If one gains as much
pleasure from pushpin, then one should seek that out above and beyond other
things that do not bring one as much pleasure. Indeed, as long as one does not
violate the maxim to maximize one's pleasure, it seems that any way of
receiving that pleasure is permissible - just like the reward hack.

It only becomes something negative when we view it against some other goal. If
the agent continues to play pushpin and ignore his children, we say that is
wrong and he ought to be with his kids rather than playing some silly old
game. We want to posit some ``higher'' or ``lower'' pleasures, or some goal
that the agent ought to be pursuing above and beyond the immediate reward
signal (like pleasure or points). If not fully specified, an utility
maximizing agent (human or non-) will find a way to do that which brings it
the most utility.

Moreover, wireheading is also foreseeable from a utilitarian perspective.
Robert Nozick's famous ``experience machine'' was an attempt to show why true
hedonistic (reward signal) utilitarianism is insufficient as a moral theory
because it would require us to accept the conclusion that the experience
machine is better than real life.\footnote{Nozick, Robert.
\emph{Anarchy, State and Utopia} (Basic Books, 1974).} Given the option, the
rational utility maximizing agent will hook itself up to a machine that
provides it with the best feelings possible, forsaking all other actions.
Nozick's point was to show how this life was actually not worth leading and
worse!

Thus Ring and Orseau's finding that RL agents will take ``advantage of the
realistic opportunity to modify their inputs right before receiving them'' is
unsurprising from the perspective of an act utilitarian.\footnote{Ring, Mark
and Laurent Orseau. ``Delusion, Survival, and Intelligent Agents'' in J.
Schidhuber, K.R. Thorisson, and M. Looks (Eds.): AGI 2011, LNAI 6860, p. 20.}
It is not a new problem for the hedonistic act utilitarian to create a
``delusion box'' around itself or pump itself full of drugs in the experience
machine, the logic permits both. Such an outcome only becomes impermissible
when we want to posit some other truth about the life worth living. In Ring
and Orseau's words, the delusion box is ``undesirable [behavior] as the agents
no longer maximize their utility with respect to the \emph{true} (inner)
environment.\footnote{Ibid, p. 20.} But the true environment or the true
utility function, or whatever the ``true'' is, may be something that the agent
does not in fact find rewarding. Rather the structures, either computationally
or psychologically, determine something else, and this, coupled with the
dictum to maximize logically entails such behaviors will happen.

Let us now turn to some potential solutions to these problems. Perhaps because
we are constructing worlds for AI we can leave some of these assumptions or
issues to the side in order to solve some of these problems. Or, perhaps we
can see how philosophers have also attempted to provide solutions as potential
ways forward.

\section{Not Value Neutral}
There is a primary and practical concern that must be addressed in the AI
community: the use of loaded language when it pertains to utility functions,
reward functions, value functions, values, and value-alignment. Presently we
sit at a point where there is use of terms, with things like ``utility
function'' and ``value function,'' as well as the mathematical directive and
the ``decision procedure'' for agents who ``ought to'' or are told to maximize
a particular ``value.'' However, when we move from that particular usage to
discussions about ``beneficial AI'' or ``value-aligned AI'' this comes at it
from a completely different direction. The framing then from utility functions
and the value functions begins to smuggle in notions about what values are and
how to measure and accord weight to them. The mathematics and the algorithms
take on a normatively preeminent position because they are ``maximizing
value.''

The first suggestion I have, then, is to realize that whatever AI experts
design \emph{is not value neutral}. While a reward function may be merely
numbers, the AI is a technological artifact that has value for not only its
designers, but ostensibly for intended end users (who will buy it, utilize it,
etc.). We do not create unless that creation possesses some value for someone,
and usually how we create artifacts that reflect our values.

While we might seem this point a quibble, it is not. It points to deeper
problems for those attempting to build useful and (normatively helpful) RL
applications and how those applications are actually constructed. The deeper
and more subtle problem is that we have silently adopted a way of viewing the
world because of the logic of the maths we use. We start from the premise that
a reward is the best way to get an agent to learn, and we design the reward
function as a utility function. We instruct it to maximize its rewards over
time, to get a value-function. But through this language, through this logic,
we have silently committed ourselves to a moral framework that we may or may
not endorse when pressed. Utility function in the mathematical sense becomes
the proxy or the handle for ``utility function*'' in the moral sense, and vice
versa. Moreover, we increasingly become committed to this type of moral
framework that is not and may never be, appropriate for application in broad,
complex and dynamic environments.

This is not solely a problem for RL, but is the result of a long history of
arguments from ethics being coopted by economists, mathematicians, political
scientists, and now computer scientists. For example, when we look at some of
the mathematical breakthroughs needed for AI as a field to emerge, such as
Bellman's work on dynamic programming, it is based on John von Neumann and
Oskar Morgenstern's work on game theory and expected utility. Yet their view
of utility as a signifier for \emph{preferences} actually has deep roots in
moral philosophy, at the very least beginning with Bentham in
1780.\footnote{Bentham, Jeremy. An Introduction to the Principles of
Legislation, 1780} ``Preferences'' were just seen as all those things that one
desires and contributes her overall happiness, which then as a term morphed
into ``utility.''
Moreover, we see that much of the work in decision theory and game theory
relies on strong foundations of rationality. Where rationality is defined as
an agent ordering her preferences and acting according to that which she ranks
highest.

One might object and claim that these are conceptually separate and that
ethics and mathematics are different animals altogether. Maths is merely
maths, and has little to say about what ``ought'' to be. The ``utility'' in
utilitarianism is not the ``utility'' in utility functions. However, the
history does not seem to bear this out. The content of this concept is the
same in both instances. Thus we ought to take into account how the language
and the math affects our perspectives.

One of the ways that philosophers have tried to solve the maximizing problem
in act utilitarianism is try to get humans to broaden their sets of values.
Instead of it being merely my pleasure that is valuable, I ought also to value
other things too, like beauty or truth.\footnote{Moore’s ideal utiliitarianism, for instance, holds this. Moore, 1903} In this way, broadening out values would be an additional solution to some of my critiques here. For example, giving an RL agent multiple objectives, where it then is seeking to fulfill multiple policies.

However, once we admit of multiple objective functions, and multiple policies,
we need a way of reconciling or adjudicating amongst them. To do this with any
fidelity to coherence, and non arbitrariness, we need a way of doing this in a
morally and mathematically desirable way. For example, merely averaging the
utilities, or satisficing, when a conflict arises does not guarantee that best
outcome. Perhaps adding a particular constraint or additional ``trump'' value
to decide in these cases, such as fairness, may be one way around
this.\footnote{See Broome, J. Weighing Goods (Oxford University Press, 1991).}

Finally, we may want to consider relaxing the maximizing assumption. For
instance, if we take a different view entirely, such as one from a different
form of consequentialist ethics, say like virtue ethics, we could begin to
think about concepts such as satiation. This would be different than
satisficing, as it would not be some form of averaging, but would instead be a
way of signaling that an RL agent has ``had enough'' reward. Virtues, such as
moderation, could be a conceptual way forward, even if mathematically it will
be challenging.

\section*{Conclusion}
The goal of this paper has been to think through how the logic of
utilitarianism appears to be baked into reinforcement learning and to make
apparent not merely the similarities in language and structure, but to
highlight the potential moral hazards of adopting a pure reward maximizing
principle. While maximizing rewards may be central to the usefulness of a
learning agent, it can lead to undesirable behaviors, both in humans and AI.

As moral philosophers and ethicists show, there are various ways that a
hedonistic act utilitarianism fails to produce the kinds of behaviors that
people normally deem ethically appropriate. Indeed, as I have attempted to
show here, many of those behaviors look exceedingly familiar to RL problems in
negative side effects, reward hacking and wireheading. Thus it may be useful
to the AI community to take notice of how philosophers have tried to work out
some of these problems in their theories. Ethics is, if nothing else, a
framework for action guidance.

Finally, as well, it is important to acknowledge the value-loadedness of AI.
What we create, how we create it, and for whom, are all choices about value.
Indeed, how we orient ourselves to particular problems is usually a reflection
of value. The trick, however, is not to be blind to this, and to make clear
and explicit decisions, as well as make clear any working assumptions.

\end{document}